%% file: paper.tex
\documentclass[letterpaper]{article}

\usepackage{aaai21}  % DO NOT CHANGE THIS
\usepackage{cuted}  % <---
\usepackage{times}  % DO NOT CHANGE THIS
\usepackage{helvet} % DO NOT CHANGE THIS
\usepackage{courier}  % DO NOT CHANGE THIS
\usepackage[hyphens]{url}  % DO NOT CHANGE THIS
\usepackage{graphicx} % DO NOT CHANGE THIS
\usepackage{natbib}  % DO NOT CHANGE THIS AND DO NOT ADD ANY OPTIONS TO IT
\usepackage{caption} % DO NOT CHANGE THIS AND DO NOT ADD ANY OPTIONS TO IT
\usepackage[draft]{hyperref}
\usepackage{graphicx,balance,multirow,xspace,subcaption,longtable}
\usepackage{makecell,amsmath,cleveref,rotating,amssymb}
\usepackage[dvipsnames]{xcolor}
\usepackage[font={small}, textfont=md]{caption}
\usepackage[utf8]{inputenc}
\usepackage[T1]{fontenc}
\usepackage[draft,inline,nomargin,index]{fixme}
\usepackage{booktabs}

\PassOptionsToPackage{hyphens}{url}
\newcolumntype{P}[1]{>{\arraybackslash}p{#1}}
\newcolumntype{X}[1]{>{\centering\arraybackslash}p{#1}}

\expandafter\def\expandafter\UrlBreaks\expandafter{\UrlBreaks%  save the current one
  \do\a\do\b\do\c\do\d\do\e\do\f\do\g\do\h\do\i\do\j%
  \do\k\do\l\do\m\do\n\do\o\do\p\do\q\do\r\do\s\do\t%
  \do\u\do\v\do\w\do\x\do\y\do\z\do\A\do\B\do\C\do\D%
  \do\E\do\F\do\G\do\H\do\I\do\J\do\K\do\L\do\M\do\N%
  \do\O\do\P\do\Q\do\R\do\S\do\T\do\U\do\V\do\W\do\X%
  \do\Y\do\Z}

\crefformat{section}{\S#2#1#3}
\crefformat{subsection}{\S#2#1#3}
\crefformat{subsubsection}{\S#2#1#3}

\fxsetup{theme=colorsig,mode=multiuser,inlineface=\itshape,envface=\itshape}
\FXRegisterAuthor{hh}{ahh}{Hussam}
\FXRegisterAuthor{mm}{amm}{Maaz}
\FXRegisterAuthor{fz}{afz}{Fareed}
\FXRegisterAuthor{rn}{arn}{Rishab}

\newcommand\clearrow{\global\let\rowmac\relax}

\newcommand{\para}[1]{{\vspace{.05in} \em \noindent #1 }}

\newcommand{\bannedsubreddit}[2]{\emph{r/#1 (banned in #2)}}

\newcommand{\subreddit}[1]{\emph{r/#1}}

\newcommand{\revred}[1]{#1}%{\textcolor{red}{\it #1}}
\newcommand{\revblue}[1]{#1}%{\textcolor{blue}{\it #1}}
\newcommand{\revgreen}[1]{#1}%{\textcolor{ForestGreen}{\it #1}}

\newenvironment{packeditemize}{
\begin{itemize}
  \setlength{\itemsep}{1pt}
  \setlength{\parskip}{0pt}
  \setlength{\parsep}{0pt}
  \setlength{\topsep}{1pt}
}{\end{itemize}}

\renewcommand{\footnotesize}{\scriptsize}
\newcommand{\etc}{etc.}
\newcommand{\eg}{e.g.,\ }
\newcommand{\etal}{et al.\xspace}
\newcommand{\ie}{i.e.,\ }

\begin{document}

\title{Are Proactive Interventions for Reddit Communities Feasible?}

\author {
    % Authors
    Hussam Habib,\textsuperscript{\rm 1}
    Maaz Bin Musa,\textsuperscript{\rm 1}
    Fareed Zaffar,\textsuperscript{\rm 2}
    Rishab Nithyanand, \textsuperscript{\rm 1} \\
}
\affiliations {
    % Affiliations
    \textsuperscript{\rm 1} University of Iowa \textsuperscript{\rm 2}, Lahore University of Management Sciences \\
    \textsuperscript{\rm 1} \{hussam-habib, maazbin-musa, rishab-nithyanand\}@uiowa.edu, \textsuperscript{\rm 2} fareed.zaffar@lums.edu.pk\\
}

\maketitle

\begin{strip} 
\begin{center} 
\vspace{-3cm}
	\textbf{This paper has been accepted at ICWSM 2022 -- please cite accordingly.}
\end{center} 
\end{strip}

\input{abstract}

\input{introduction}
\input{background}

\input{evolution}

\input{predictors}

\input{relatedwork}

\input{discussion}

\balance

%\small

\bibliographystyle{aaai21}
\bibliography{reddit}

%\newpage

%\input{appendix}

\end{document}

%% file: abstract.tex
\begin{abstract}
  Reddit has found its communities playing a prominent role in
  originating and propagating problematic socio-political
  discourse. Reddit administrators have generally struggled to prevent or
  contain such discourse for several reasons including: (1) the inability for
  a handful of human administrators to track and react to millions of posts and
  comments per day and (2) fear of backlash as a consequence of administrative
  decisions to ban or quarantine hateful communities. Consequently,
  administrative actions (community bans and quarantines) are often taken only
  when problematic discourse within a community spills over into the real world
  with serious consequences.  
  In this paper, we investigate the feasibility of deploying tools to
  proactively identify problematic communities on Reddit. Proactive
  identification strategies show promise for three reasons: (1) they
  have potential to reduce the manual efforts required to track communities for
  problematic content, (2) they give administrators a scientific rationale to
  back their decisions and interventions, and (3) they facilitate early and
  more nuanced interventions (than banning or quarantining) to
  mitigate problematic discourse. 
  
%  Our work shows that communities are constantly evolving in
%  their user base and topics of discourse. This suggests that human-only
%  administration faces challenges of scalability. Further analysis reveals that
%  evolution into problematic (\ie considered bannable by Reddit administrators)
%  communities can often be predicted months ahead of time. This suggests that
%  proactive administrative strategies are feasible. We also leverage
%  explainable machine learning to help identify the strongest predictors of
%  evolution into problematic communities. While our analysis provides an
%  understanding of the biases currently at play in Reddit's administrative
%  decisions, it additionally provides administrators with insights into the
%  characteristics of communities at risk of becoming problematic in the future.
%   Finally, we investigate, at scale, the effectiveness of current
%   administrative interventions such as bans and quarantines on the behavior of
%   members of problematic communities. Our findings here highlight the
%   weaknesses of such interventions when they are inconsistently applied and the
%   need for more nuanced interventions -- particularly when working with
%   proactive administrative strategies.
\end{abstract}

%% file: introduction.tex
\section{Introduction}\label{sec:introduction}

% \para{Controversies on the ``front page of the Internet''.}
% Opening statement
Reddit has over 138K active communities, called \emph{subreddits}, with over
330M  active users making it the 6th most popular website in the
USA.%\cite{Reddit-Web2019}.  % Current knowledge and bait
In recent years, the site has been mired in controversies around the
role that its communities played in originating and propagating sexist,
racist, and generally hateful online socio-political discourse. A few of the
recent controversies have involved communities such as:
\bannedsubreddit{Physical\_Removal}{8/2017} which advocated for the physical
removal of `liberals' in the United States prior to and even after the murder of
Heather Heyer in Charlottesville \cite{HeatherHeyer-DailyBeast2017},
\bannedsubreddit{incels}{11/2017} which endorsed and celebrated the murder of
and violence against sexually active women \cite{AlekMinassian-NBCNews2018},
and \bannedsubreddit{greatawakening}{3/2018} and
\bannedsubreddit{pizzagate}{11/2016} which falsely alleged the existence of
child trafficking rings by the US Democratic Party and left-wing corporations
resulting in real-life attacks, threats, and harrassment \cite{Pizzagate-WaPo2016}.
%\bannedsubreddit{WatchPeopleDie}{3/2019}, \bannedsubreddit{gore}{3/2019}
%which disseminated videos of the Christchurch Mosque shootings
%\cite{Christchurch-TheVerge2019}, and \emph{The\_Donald (quarantined in
%6/2019)} which continues to pedal white genocide conspiracy theories
%\cite{TheDonald-WhiteGenocide-SPLC2018}, false-flag theories in the wake of
%shootings and bomb threats \cite{TheDonald-FalseFlags-NYT2018}, and violent
%anti-immigrant \cite{TheDonald-AntiImmigrant-FC2018} and anti-Islam rhetoric
%\cite{TheDonald-AntiIslam-MotherJones2019}. 

%\para{The challenges of reactionary moderation strategies.}
% Setting up the gap in knowledge
In reaction to many of these controversies, Reddit has resorted to banning or
quarantining subreddits citing violations of the Reddit content policy
% \footnote{\url{https://www.redditinc.com/policies/content-policy}}
which prohibits specific types of content
including content which ``\emph{encourages or incites violence}''.
However, the effectiveness and timeliness of such bans and quarantines are
frequently debated. While previous research \cite{Chandrasekharan-CSCW2017}
concluded that such bans ``\emph{worked for Reddit}'', others have pointed out
that they are too reactionary and occur only after a significant amount of
damage has already been observed \cite{RedditModeration-Mashable2019,
RedditModeration-Vox2017, Marantz-NY2018}. Along another dimension, Reddit has
also faced criticism for inconsistent and seemingly ad-hoc applications of the
content policy by those claiming that the platform provides a safe-haven for
extremist ideologies and others claiming that the platform leverages the
content policy as a mechanism to censor ``non-mainstream'' opinions and
ideologies. Furthermore, Reddit admins and moderators have claimed that the
non-static nature of communities requires them to perform constant monitoring
and community guidelines updates \cite{seering2019moderator} -- a task which
makes administration more challenging.
Despite these criticisms and known challenges surrounding Reddit
administration, little is actually known about the evolutionary characteristics
of subreddits and the predictors of problematic subreddits (\ie \emph{those
deemed to have violated the content policy}). Further, there are no publicly
available tools to help Reddit administrators make timely and sound
intervention decisions. We seek to fill these gaps by studying the evolutionary
characteristics of subreddits and developing an administrative tool to help
with early identification of potentially problematic subreddits. This report
describes our analysis and methods related to the following two hypotheses.

  \para{H1. Subreddits do not converge to stability.} 
  \revgreen{This hypothesis demonstrates the need for automated
  tools to monitor subreddit evolution. If valid, it shows that constant
  monitoring is required for subreddits due to the evolving nature of discourse
  and participation -- a prohibitively expensive proposition for administrators
  without automated monitoring tools.} In order to test this hypothesis, we
  develop techniques to track subreddit evolution in terms of vocabulary and
  participating users. Our work validates this hypothesis.
  
  \para{H2. Evolution in problematic subreddits can be predicted.} This
  hypothesis demonstrates the promise of automated tools to help administrators
  make timely and sound intervention decisions. If valid, it shows that
  pre-emptive identification of subreddits likely to violate Reddit's content
  policy is possible. In order to test this hypothesis, we develop explainable
  ML models which rely on a variety of features including structural-,
  linguistic-, community-, and user-related features to predict the
  evolutionary outcome of a subreddit. Our models show that that problematic
  communities can be identified by their evolutionary behavior early in their
  lifetime. Further, the explainability of our models provides administrators
  with an understanding of the causes for classification decisions and the
  ability to use their expertise to overrule and augment them. Finally, we
  deploy our classifier in a real world continuous learning scenario, identical
  to its use-case   for Reddit, and study its predictions. 
  
  Taken together, our study demonstrates the feasibility of proactive and
  explainable machine-aided strategies to help, but not replace, human
  administration of Reddit. 

%% file: background.tex
\section{Reddit: The Platform and Dataset}\label{sec:background}

%Reddit, also known as the `\emph{front page of the internet}'
%\textsc{\color{green} using this line 4th time is chessy no?}, is a social
%media and a content aggregation site. 
%Reddit is currently the sixth most popular site in the USA with over 330M
%active monthly users \cite{Reddit-Web2019}.

% In recent years, Reddit has come under increasing criticism for the types of
% content shared by its users \cite{TheDonald-WhiteGenocide-SPLC2018,
% TheDonald-FalseFlags-NYT2018, TheDonald-AntiImmigrant-FC2018,
% TheDonald-AntiIslam-MotherJones2019}. This has resulted in numerous changes of
% the Reddit content policy and administrative strategies. 
In this section, we
provide a high-level overview of Reddit with a focus on its content and
administrative policies (\Cref{sec:background:overview}) and the
datasets we rely on (\Cref{sec:background:data}).

\subsection{An overview of the Reddit platform}\label{sec:background:overview}
% \para{Redditors and subreddits.} At a high-level, Reddit is a content
% aggregation platform where users, also called \emph{redditors}, share content
% on topical forums called \emph{subreddits}. 
% % Subreddits are generally formed around specific topics and can range from
% % broad (\eg \subreddit{politics} which focuses on US politics) to extremely
% % niche (\eg \subreddit{birdswitharms} which focuses on photoshopped images of
% % birds with human arms). 
% There are currently 138K active subreddits \cite{Reddit-Web2019}. In addition
% to posting content, redditors can also interact with each other by commenting
% on posts and replying to other comments.

% \para{Democratized content-curation.} Unlike other social and aggregation
% platforms, Reddit relies on its users for more than content generation and
% propagation. Redditors also play the role of content curators. Redditors may
% curate content by up- or down-voting comments and posts. Content (comments
% or posts) with a high net-vote total is, by default, given very high
% visibility. For example, the comments (on a post) and posts (in a subreddit)
% are, by default, ordered by decreasing net-vote total. This mechanism lets
% users decide which comments and posts are most visible to the rest of the
% community and which contributions are silenced or hidden. 

%\para{Decentralized moderation.} 
% Besides letting every redditor curate
% content by way of voting, 
Reddit allows its users to create and moderate
subreddits. Subreddit moderators typically choose their own fellow moderators
from within the community, with a few exceptions for newly created communities
and cases where there are no volunteers within the community. %In these cases,
%the subreddit creator is made the moderator or these moderator positions may be
%filled through other means such as through \subreddit{adoptareddit} and
%\subreddit{needamod}. 
Subreddit moderators are tasked with setting and enforcing the rules of
engagement within a subreddit. Moderators may enforce rules via the use of user
bans and content deletion. However, the actions of subreddit
moderators do not impact redditor experiences outside of that subreddit (\eg
a subreddit moderator cannot enforce site-wide bans). 
%\para{Centralized administration.} 
In addition to relying on volunteers, Reddit
also employs administrators to set and
enforce site-wide policies for content and user engagement. These content
policies are
mandatory and applied \emph{in addition to} a subreddit's own policies.
In the event of content policy violations, administrators have the ability to:
(1)  ban users from making posts or comments visible to the rest of the
platform (\ie \emph{shadow ban}), (2) prevent subreddits from appearing on the
Reddit front-page and in search results (\ie \emph{quarantine}), and (3) ban
subreddits from the platform. Currently, this administration process is largely
manual requiring a team of administrators to manually study reports of content
violations submitted by Redditors with little support provided in the way of
automated decision-making aids.
Beyond poor scalability, these manual efforts have also impacted the mental
health of content-policy administrators on Reddit \cite{We-are-the-nerds}.

\subsection{Datasets}\label{sec:background:data}
In this paper, we focus on specific subsets of the entire platform. These
were gathered using the Pushshift API \cite{pushshift}. \revred{We note that due to
computational limitations, the vocabulary vectors used in our analysis
(described in \S\ref{sec:evolution:methods:vectors}) leverages random 10\%
samples of each of the datasets below. However, the remainder of our analysis
has no such limitations.}

\para{Most active subreddits ($\mathfrak{D}_A$).} \revred{This dataset contains
the 424M posts and 4.5B comments made by 42M unique users to the 3K most
active subreddits (\ie highest average number of comments per month) which did
not receive any administrative interventions (\ie bans or quarantines) during
the period from 01/2015 to 04/2020.}

\para{Subreddits with administrative interventions ($\mathfrak{D}_I$).} \revred{This
dataset contains 38M posts and 353M comments made by 2.3M unique users from
264 subreddits which, between 2015 to 2020, were the subject of either
administrative bans,  quarantines, or both.} Since most bans or quarantines are
not announced by the Reddit administrative team, the list of banned subreddits
was obtained by visiting the webpages of subreddits in which user activity had
ceased and confirming the presence of a ban notification from Reddit
\footnote{see \url{https://www.reddit.com/r/incels/} for an example}. The date
of the last post made on the subreddit was used as the ban date for each banned
subreddit. To identify quarantined subreddits, we leveraged data from
\subreddit{reclassified} which lists a crowd-sourced subset of all quarantine
events on the Reddit platform. Quarantine actions were confirmed by visiting
the webpages of the subreddits and confirming the presence of a quarantine
notification from Reddit. The date of the post on \subreddit{reclassified} was
used as the quarantine date for each quarantined subreddit. 
Since our goal is solely to capture the characteristics of
problematic communities, we consider bans and quarantines as equivalent.

\para{Control subreddits without administrative interventions
($\mathfrak{D}_C$).} Since $\mathfrak{D}_A$ and $\mathfrak{D}_I$ have vastly
different sizes and contain subreddits with different amount of activity, we
create a control dataset to facilitate comparisons with $\mathfrak{D}_I$. This
dataset contains subreddits most similar to those in $\mathfrak{D}_I$ along
two parameters: vocabulary and activity. For each subreddit in $\mathfrak{D}_I$ and
$\mathfrak{D}_A$, we create two vectors: a vocabulary vector (using the techniques
outlined in \Cref{sec:evolution:methods}) and an activity vector which denotes
the number of comments in the subreddit during each month. For each subreddit in
$\mathfrak{D}_I$, we then find the subreddit in $\mathfrak{D}_A$ which has the
most similar topic and activity vector. Similarity is computed by cosine
similarity and weights are equally assigned for the topic and activity vector.
We manually verified the similarities of each matched pair of subreddits.
Examples of $\mathfrak{D}_C$ subreddits and their corresponding
$\mathfrak{D}_I$ subreddits are: (\subreddit{conspiracy},
\subreddit{911truth}), (\subreddit{Conservative}, \subreddit{The\_Donald}),
(\subreddit{PurplePillDebate}, \subreddit{MGTOW}), and (\subreddit{niceguys},
\subreddit{Incels}). \revred{This dataset contains 44M posts and 489M comments
made by 44M unique users to all $\mathfrak{D}_I$-matched subreddits.}

%% file: evolution.tex
\section{Subreddit Evolution and Convergence}\label{sec:evolution}

In this section, we focus on testing the following hypothesis:  \textbf{H1.
Subreddits may not converge to stability}. We measure stability by the
vocabulary of the community and the `backgrounds' of the users participating
them them. If valid, this hypothesis demonstrates the need: (1) for
computational techniques to monitor subreddit evolution (in terms of vocabulary
and user bases) and (2) to frequently evaluate the suitability of making
administrative interventions. \revgreen{Put another way, the validity of this
hypothesis would suggest vocabulary and user base in communities are always
evolving and therefore require consistent and frequent monitoring --- a task
which is known to be non-scalable, expensive, and extremely laborious for human
moderators \cite{We-are-the-nerds, roberts2014behind,
wohn2019volunteer}}.
% Additionally,
% such analysis may yield insights into the potential for identifying
% evolutionary patterns that can serve as early predictors for subreddits likely
% to require future administrator interventions. If invalid, the failed
% hypothesis will show that subreddits converge to topical or user base stability
% and therefore only need to be evaluated by administrators once -- after
% stability has been reached. This will invalidate the need for computational
% tools since the constant monitoring of subreddits is unnecessary. Our methods
% and results are outlined in \Cref{tab:evolution}.
In order to test this hypothesis, we conduct analysis to check if the
vocabulary used and users participating in subreddits in $\mathfrak{D}_A$,
$\mathfrak{D}_I$, and $\mathfrak{D}_C$ stabilize over time.  
\subsection{Methods}\label{sec:evolution:methods}
% Our method for testing the hypothesis involves two steps: First, we create
% latent vector summaries of topics and user participation for each subreddit
% state (\Cref{sec:evolution:methods:lsa}). Next, we measure the similarities of
% these summaries as a function of time (\Cref{sec:evolution:methods:clustering}).
% 
\subsubsection{Representing subreddit vocabulary and users with fixed-length
vectors.}\label{sec:evolution:methods:vectors}

A \emph{subreddit state} is all the activity associated with
a subreddit during a given month -- \eg the subreddit state associated with
\subreddit{politics} on 09/2019 contains all user posts and comments made on
\subreddit{politics} during the month of 09/2019. Our goal is to create two
fixed-length vectors which capture: (1) the vocabulary associated with each
subreddit state and (2) the active user base associated with each
subreddit state.

\para{Creating fixed-length vocabulary vectors for each subreddit state.} At
a high-level, our vocabulary vector for each subreddit state is the vector of TF-IDF
weights associated with each unique token in our dataset. 
% This is a standard document
% representation approach in the information retrieval community
% \cite{Beel-IJDL2016} to associate keywords with documents. 
% We apply this approach as follows:
\begin{packeditemize}
  \item \textit{Random sampling and document corpus creation.} \revred{We begin by
    randomly sampling 10\% of all comments in our dataset (a necessity owing to
    the large dataset and computational limitations).} The sampled comments are
    then used to create documents associated with each subreddit state -- \eg
    the document associated with the (\subreddit{news}, 09/2019) subreddit
    state will contain all the sampled comments which were made by users on
    \subreddit{news} during the month of 09/2019. At the end of this step, we
    have a corpus of documents ($D$) containing one document for each
    subreddit state in our dataset. 
  \item \textit{Text pre-processing and token corpus creation.} We perform
    standard text pre-processing operations including removing English stop
    words, tokenization, and lemmatization for each document. The unique
    tokens, across $D$, at the end of this pre-processing stage form the corpus
    of words and determine the length of the vectors associated with each
    subreddit state. At the end of this step, we have a corpus of all unique
    lemmatized tokens ($T$) observed in $D$. The length of the topic vector
    associated with each subreddit state is $|T|$.
  \item \textit{Computing vocabulary vectors for each subreddit state.} For each
    document $d \in D$, we compute the TF-IDF weight of each token $t \in T$.
    Therefore, the vocabulary associated with
    each subreddit state are represented by a vector denoting the importance of
    each token with respect to all sampled comments made on
    the subreddit during the corresponding month.
\end{packeditemize}
    % Therefore the $i^{th}$ entry in the topic vector for $d$ is
    % ${f(t_i)}\times{\log\frac{|D|}{|D_{t_i}|}}$, where $f(t_i)$
    % denotes the frequency of token $t_i$ in $d$ and $D_{t_i}$ is the set of all
    % documents containing the token $t_i$. A high TF-IDF weight for a (document,
    % token) pair indicates that the token has a high frequency of occurrence in
    % the given subreddit state document and a low frequency of occurrence in all
    % other subreddit state documents -- \ie the token is of high importance in
    % the given subreddit state document. 

\para{Creating fixed-length active user vectors for each subreddit state.} At
a high-level, our active user vector for each subreddit state is the vector of
fractions of active user co-occurrences with other subreddit states. This is
a standard technique used in collaborative filtering and recommender systems
research \cite{Liang-RecSys2016, Dunning-2014} to identify shared interests
between sets of users.% or communities. % We apply this approach as follows:
\begin{packeditemize}
  \item \textit{Active user identification.} We begin by creating sets of
    active users for each subreddit state in our dataset. Active users
    associated with a subreddit state are identified as all users with any
    posting or commenting activity on the corresponding subreddit during the
    corresponding month. We refer to the set of active users of subreddit state
    $s$ as $A_s$. Active user vectors were created using the complete dataset
    without sampling.
  \item \textit{Computing active user vectors for each subreddit state.} For
    each subreddit state, we compute the fraction of active users who overlap
    with every other subreddit state. The $j^{th}$ entry in the active user
    vector for subreddit state $i$ corresponds to the fractional overlap with
    subreddit state $j$ -- \ie it has the value $\frac{|A_i \cap
    A_j|}{|A_i|}$. A higher overlap ratio between two subreddit states
    indicates that the two subreddits shared a large cohort of users during the
    specific month. %Therefore, the active users of
    %a subreddit state are represented by a vector denoting the sizes of the
    %cohorts shared with all other subreddit states. 
    The length of this vector
    is equal to the total number of subreddit states in our dataset.
\end{packeditemize}

\subsubsection{Measuring the distance between two subreddit states.} 
\label{sec:evolution:methods:clustering}
% Here, in contrast with absolute distance measures, our goal is to define the
% distance between two subreddit states as a function of all other subreddit
% states in our dataset. In other words, our measure needs to track subreddit
% evolution while accounting for the effects of changes occurring on other
% subreddit states.
% 
% \para{Relative distance measures and their relationship with subreddit
% evolution.}
There are two types of distance measures that are available for use: absolute
(\eg cosine similarity between the vectors of two subreddit states) and
relative (\eg similarity of ranked lists of nearest subreddit state neighbors
for two subreddit states).  We chose to leverage the latter. Note, however,
that our analysis with absolute distance measures yielded similar results to
those presented in \Cref{sec:evolution:results}.
Relative distance measures, in the context of subreddit states,
allow us to quantify how a subreddit's vocabulary and user bases have evolved as
a function of the vocabularies and user bases of other subreddits. For example, let
us once again consider the states associated with (\subreddit{news}, 11/2019)
and (\subreddit{news}, 04/2020). The relative distance between the vocabulary
vectors associated with these subreddits will account for the fact that
although the absolute change between the vectors is large due to the change in
discussion topics from American primary elections to COVID-19, the changes with
relative to other subreddits is smaller -- \ie subreddits (\eg
\subreddit{usnews}) which shared similar topics with \subreddit{news} in
11/2019 still shared similar topics in 04/2020. Analysis with a relative
distance measure therefore identifies how much subreddits have changed with
respect to their neighbors and how their role on the platform has changed. We
say that convergence has occurred if the relative distance computed over
consecutive months converges to the minimum.

\para{Quantifying relative distance using Rank-Biased Overlap (RBO)
\cite{Webber-IS2010}.} Given two vectors of identical length which represent
either vocabulary or active user base vectors of two subreddit states, we perform
the following operations to obtain their relative distance.

\begin{packeditemize}

\item \textit{Generate list of neighbors ordered by euclidean distances.} Let
    $v_1$ and $v_2$ be the two input (topic or user base) vectors, associated with
    subreddit states $s_1$ and $s_2$ and belonging to months $m_1$ and $m_2$, whose
    relative distance we wish to compute. Let $S_{m_1}$ be the set of all subreddit
    states from month $m_1$ and $S_{m_2}$ be the list of all subreddit states from
    $m_2$. We begin by computing the euclidean distances between (1) $v_1$ and every
    subreddit state in $S_{m_1}$ and (2) $v_2$ and every subreddit state in
    $S_{m_2}$. Finally, we sort the elements of $S_{m_1}$ and $S_{m_2}$ in ascending
    order of their euclidean distance to $v_1$ and $v_2$ and store the sorted list
    in $X_1$ and $X_2$, respectively. These lists are effectively the subreddit
    neighbors of $s_1$ and $s_2$ during $m_1$ and $m_2$, respectively.

\item \textit{Computing RBO scores.} Given the lists of neighbors, $X_1$
  and $X_2$, we then compute the RBO similarity score between them. We use RBO
    since it automatically imposes higher penalties for disagreements at top
    ranks and works for non-conjoint and arbitrarily long ranked lists. These
    properties are not available using methods such as Kendall's Tau rank
    similarity metrics. A high similarity score ($\approx 1$) indicates a low
    relative distance between the two subreddit states -- \ie the two states
    have nearly identical sets of nearest neighbors. 
\end{packeditemize}

\subsection{Results}\label{sec:evolution:results}

\begin{figure} \centering 
   \includegraphics[width=.4\textwidth]{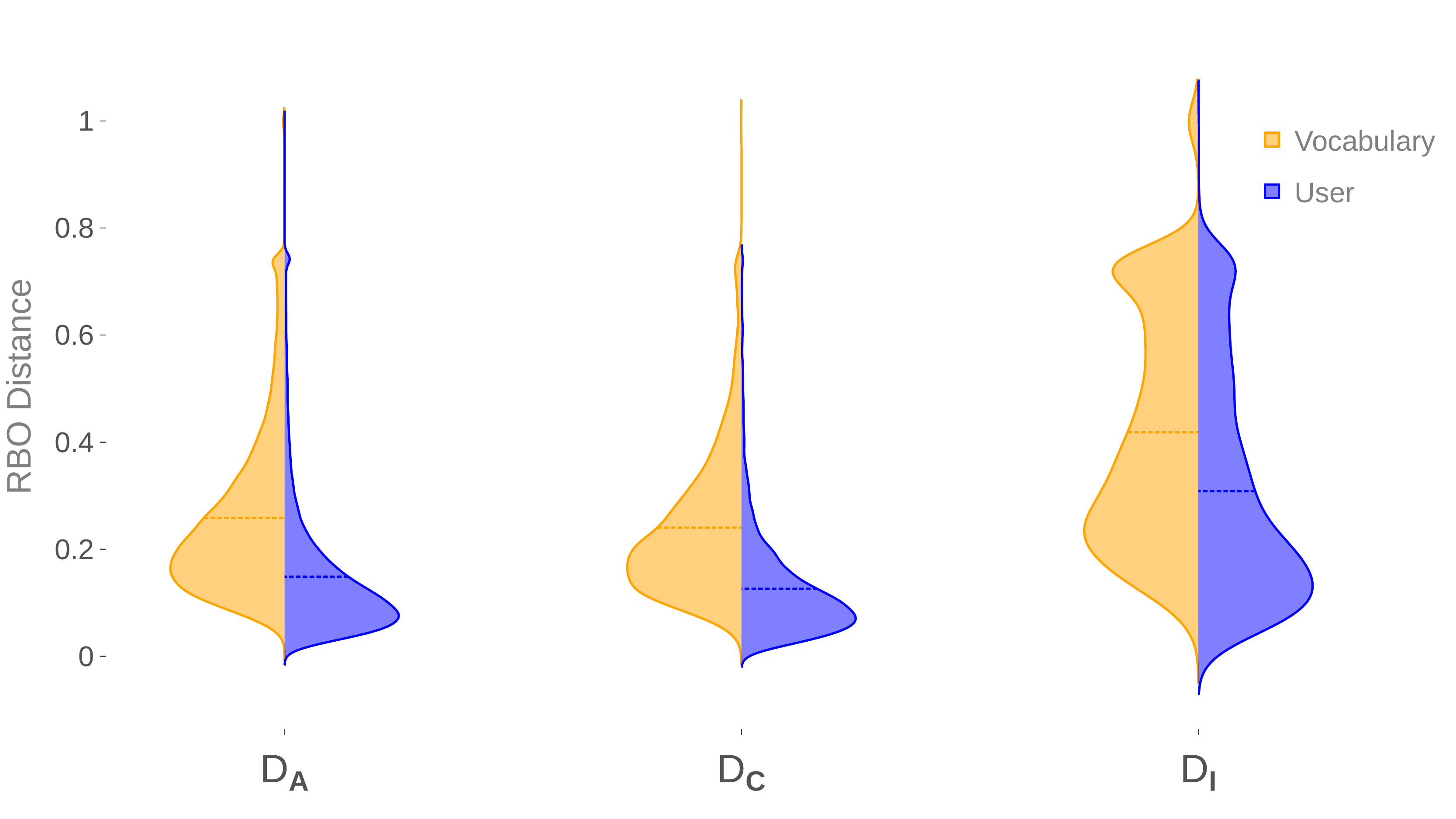}
   \caption{Distribution of RBO distances of vocabulary/user vectors
   between consecutive months for subreddits in $\mathfrak{D}_{A}$,
   $\mathfrak{D}_{C}$, and $\mathfrak{D}_{I}$.}
  \label{fig:evolution:results} 
\end{figure}

% We now highlight the results of our analysis of how subreddit topics and user
% bases evolve over time. In our analysis we focused on tracking the absolute and
% relative distance measures to understand how topics and user bases evolved for
% subreddits from $\mathfrak{D}_I$ and $\mathfrak{D}_A$. We restrict our analysis
% to the first {48} months of each subreddit's lifespan.
% during the first four years of their lifespan using two basis for comparison:
% (1) the first month of their lifespan (\ie we compute the distance between
% vectors from months 1-48 with month 0) and (2) the prior month (\ie we compute
% the distance between month $i$ and month $i-1$ for each of months 1-48).

\Cref{fig:evolution:results} shows the results of our measurement of
monthly subreddit vocabulary evolution using a relative distance measure
(RBO). The plot shows the distribution of the measured RBO distances between
any two consecutive months for each subreddit in $\mathfrak{D}_A$,
$\mathfrak{D}_C$, and $\mathfrak{D}_I$. We can make several observations from
these results. First, it appears that, on average, there is a consistent change
in vocabulary from month to month -- regardless of the subreddit category. We
see smaller changes in the active user cohorts from one month to the next, on
average, however. Second, there is a statistically significant difference (KS
test, $p$ < .01) between the monthly (vocabulary and active user cohort)
changes seen by subreddits in $\mathfrak{D}_I$ and all other subreddits in our
dataset. This is evident by the observed bimodal RBO distance distribution seen in
$\mathfrak{D}_I$. \revblue{To ensure the consistency of our results related to
the evolution of vocabulary vectors (which were gathered on a 10\% sample of
our dataset), we repeated our analysis on three independent 10\% samples and
confirmed the statistically significant differences between the RBO distance
distributions between consecutive months for subreddits in $\mathfrak{D}_I$ and
all other subreddits in our dataset.} Further,
manual validation confirmed that the months showing higher RBO 
distance to the prior months were the result of abnormal activities.  For
example, the \subreddit{The\_Donald} subreddit observed anomalous evolution of
active user cohort in late 2015 when a migration of active users from
\subreddit{european},  an extremist subreddit which was eventually quarantined
and banned by Reddit, was observed. Other large migrations appear to occur on
$\mathfrak{D}_I$ subreddits throughout their lifespan. One hypothesis is that,
similar to the above outlined case of \subreddit{european}, eventually
problematic subreddits see large migration events when currently problematic
subreddits are  banned. This hypothesis is supported by the results from a
previous study \cite{ribeiro2020does}. Put another way, active users of a banned
community migrate to a new community which eventually sees the same
administrative action imposed on it due to the eventual occurrence of the same
problematic behaviors.  Studying the largest changes in the $\mathfrak{D}_A$
subreddits, we see that \subreddit{feminism} had an RBO vocabulary distance of
.65 when comparing 2016/10 and 2016/11. Closer inspection shows that the
vocabulary change is largely driven by Hillary Clinton's loss in the 2016 US
Presidential elections.  Other prominent examples of $\mathfrak{D}_A$ subreddits
with large changes were: the active user cohort for \subreddit{Australia} during
01/2020 which corresponds to the outbreak of the wildfires and the active
user/vocabulary of \subreddit{newzealand} in 02/2019 following the Christchurch
Mosque shootings.  As a consequence of these real-world events, both subreddits
saw increased activity from redditors not usually active on the subreddits.
These observations suggest that the large changes in $\mathfrak{D}_A$ subreddits
are often driven by external events while changes in $\mathfrak{D}_I$ subreddits
are largely driven by on-platform administrative actions and community raids.

\para{Takeaways.} Our analysis confirms our hypothesis: {\it Subreddits may not
converge to a stable vocabulary or user base.} On average, subreddit vocabulary
evolves at a higher rate than subreddit active user cohorts. Interestingly, we see that the monthly changes observed
by $\mathfrak{D}_I$ subreddits are, on average, statistically significantly
higher than when compared to all other subreddits.  This suggests that the
differences in evolutionary patterns as well as an understanding of the causes
for large changes (\eg were they due to on-platform or real world events?) might
allow for early detection of potentially problematic subreddits. For example,
tools which identify the subreddits which are the targets of mass migrations
from recently banned subreddits might facilitate early interventions to prevent
degradation of the target community.  We operationalize these insights in
\Cref{sec:predictors}.

%% file: predictors.tex
\section{Identifying Predictors}\label{sec:predictors}

In this section, we test the following hypothesis:
\textbf{H2. Evolution into problematic subreddits can be predicted}. If valid,
this hypothesis will show that tools may be built to help moderators
pre-emptively identify subreddits likely to devolve into problematic
subreddits. In order to test this hypothesis, we extract a variety of features
from different points in a subreddit's lifespan and utilize explainable ML to
understand the predictive capabilities of each of these features and perform
early identification of problematic communities. We then evaluate the
performance of our explainable models in a real world continuous learning
setting, similar to how Reddit administrators may leverage them.

\subsection{Methods}\label{sec:predictors:methods}

\subsubsection{Extracting subreddit features.}
\label{sec:predictors:methods:features}

For each subreddit in our datasets, we break their lifespan into four quarters
and extract features from each. With this approach: (1) we are able to get an
identical number of features from all subreddits -- even if they have vastly
different lifespan values, and (2) we are able to capture features from
different phases in the evolution of a subreddit. Our extracted features fall
in six categories (listed in \Cref{tab:predictors:features}): community-,
moderator-, user-, structure-, mentions, language-, and vocabulary-related
features. These features were largely influenced by the insights from our
analysis in \Cref{sec:evolution} and existing literature seeking to predict
community dynamics (\Cref{sec:related}).

\para{Community features.} This category of features captures the dynamics of
the interactions occurring within the community -- \eg how large is the active
community, how highly do community members rate each others posts, \etc

\para{Moderator features.} Moderators play a large role in directing the growth
and policies within each community. This category of features captures how the
moderator team interacts with the community -- \eg how many moderators
does the community has, how many comments are removed by moderators, \etc 

\para{User features.} This category of features captures characteristics of the
average user within a community -- \eg how active are users, how frequently
do they delete their comments, \etc 

\para{Structural features.} We introduce a category of features to capture how
a subreddit is connected (in terms of shared user base) to other communities --
\eg how isolated is the subreddit, what fraction of its connections are to
other communities which were previously classified as problematic, \etc 

\para{Mention features.} This category of features represent the
mentions of a subreddit on other subreddits and in popular media. To obtain
these features, we identified the number of: (1) news articles written about
the specific subreddit prior to the end of the quarter being studied. The
dates and article counts were obtained using MIT and Harvard's Media Cloud
\footnote{\url{https://mediacloud.org/}}, an open source platform that gathers and
tracks content of online news, with the search restricted to their U.S. Top
Online News collection, and (2) references to the specific subreddit on
comments made on other subreddits. Finally, for both types of mentions we
compute the sentiment towards the community and categorize the mention as
either negative or not.

\para{Language features.} This category represents the language
style of the users in these communities. We use LIWC 2015
\cite{LIWC} to extract language style features. These features help
understand the psychometrics of the language within the community. In 
addition, we use the Perspective API
\footnote{\url{https://www.perspectiveapi.com/}}, a toxic speech classifier
developed by Google, to identify toxic comments. We also include the similarity
of the community's vocabulary vector with the vectors of previously known
problematic communities.

\begin{table}[t]
\small
  \begin{tabular}{lp{2.35in}}
	\toprule
	{\textbf{Category}}  & {\textbf{Features}}\\
	\hline
    \textbf{Community} & \# active unique commenters, \# posts, \# comments,
    dist. of comments \& posts, dist. of score \& comments, \% of active user
    growth, dist. of controversial score per post, \# controversial comments,
    \# gilded comments			\\
    
    \midrule
    
    \textbf{Moderators} & \# moderators, \# incoming moderators, \# outgoing
    moderators, dist. of moderators comment score, \# AutoModerator comments,
    dist. of moderators comment score, \# removed posts, \# removed posts score                	\\

    \midrule

    \textbf{Users} & \# active months, \# comments by deleted accounts	\\

    \midrule

    \textbf{Structural} & \# uniquely connected communities, \# total
    connections, \# users with connections to	previously banned communities		\\

    \midrule

    \textbf{Mentions} & \# of mentions in other communities \# of negative
    mentions in other communities, \# of mentions in popular news outlets, \#
    of negative mentions in popular news outlets	\\

    \midrule

    \textbf{Language} & 93 standard LIWC 2015 features, \# toxic comments, mean
    cosine similarity between vocabulary vectors of previously banned
    communities	\\

\bottomrule
\end{tabular}
 \caption{Features extracted from each quarter of a subreddit's life.}
  \vspace{-.25in}
 \label{tab:predictors:features}
\end{table}

\subsubsection{Preventing feature ``leakage''.}
\label{sec:predictors:methods:leakage}

Extracting features without careful considerations can result in leakage that
impacts the quality of the classification task. To avoid this problem we need
to make sure that features used in our task are actually available for use, by
administrators, at the time of the classification task. For example in Q1 of
a subreddit's lifetime which may be between months $m_1$ and $m_2$, we cannot
extract features which might rely on data from after $m_2$ -- from the
community or from external communities. This is
particularly important when considering the features in the structural,
mentions, and vocabulary categories. In our feature extraction process for
each subreddit, we take care to only consider information available from
each quarter. For example, when extracting the `\% of users with
connections to previously banned communities' structural feature we only
consider connections with communities which were banned {before the
end of the corresponding quarter, \ie $m_2$, for the corresponding
subreddit}. Similarly, when creating vocabulary vectors, the vocabulary is
limited to tokens observed only prior to $m_2$. This is maintained for every
subreddit active during $m_2$. \emph{Therefore, our features
are all obtained from data that is available to administrators at the
time of classification and are leakage-free}. %The
%classification performance results shown in \Cref{sec:predictors:results}
%are therefore representative of the effectiveness of the classifier \emph{only
%given the information available to administrators at the time of decision
%making}.

\subsubsection{Accounting for class label imbalances in training and testing.}
\label{sec:predictors:methods:skew}

Our dataset has a severe imbalance of class labels with only 264
$\mathfrak{D}_I$ subreddits and 3K $\mathfrak{D}_A$ subreddits. We take care
to address the model-building and performance-reporting challenges that arise
from this imbalance. We used and evaluated
two standard approaches for model training in the presence of imbalanced class
labels -- oversampling from the minority class ($\mathfrak{D}_I$) using
ADASYN \cite{he2008adasyn} and undersampling from the majority class
($\mathfrak{D}_A$) using ensemble learning \cite{Zhu-CDL2013,
Liu-SMC2009}.

%\begin{packeditemize}

\para{Approach 1: Oversampling $\mathfrak{D}_I$.} ADASYN
oversampling works by creating synthetic samples from the minority
class. ADASYN is very similar to other synthetic oversampling techniques
such as the SMOTE algorithm \cite{Chawla-IJCAI2002}. However, unlike
SMOTE, ADASYN adaptively creates synthetic points while considering the
neighborhoods of the class borders. Consequently, it generates synthetic points
near the class borders and mitigates the challenges associated with overfitting
(\eg seen in SMOTE oversampling). For the purposes of training our model, we
used ADASYN to create a perfectly balanced dataset with equal
numbers of $\mathfrak{D}_I$ and $\mathfrak{D}_A$. Validation and testing
were carried out only on samples not used or generated by ADASYN.

\para{Approach 2: Undersampling $\mathfrak{D}_A$.} Our
undersampling approach works by: (1) splitting the training samples of the
majority class ($\mathfrak{D}_A$) into equal sized datasets
($\mathfrak{D}_{A_1}$, $\mathfrak{D}_{A_2}$, \dots, $\mathfrak{D}_{A_n}$)
with the number of samples in each dataset equal to the total number of
minority class ($\mathfrak{D}_I$) training samples, (2) training a set of
classifiers, $c_1$ \dots $c_n$, with each using one of newly split
majority class datasets and the entire minority class training samples as
training input -- \ie classifier $c_i$ trains on samples from
$\mathfrak{D}_{A_i}$ and $\mathfrak{D}_I$, and (3) assigning the label
output by the majority of the $n$ classifiers when given a feature set for
classification into $\mathfrak{D}_I$ or $\mathfrak{D}_A$.

%\end{packeditemize}

Results for both sampling approaches are comparable and reported in
\Cref{sec:predictors:results}. In addition to the aforementioned sampling
techniques we repeat our experiments using our control dataset
($\mathfrak{D}_C$), SMOTE \cite{Chawla-IJCAI2002}, BorderlineSMOTE
\cite{han2005borderline}, and random oversampling. Using
these additional techniques yielded similar results. We note that the sampling
approaches were utilized \emph{only for expanding the training sets used
and did not impact the testing and holdout datasets} in our classification
experiments. To avoid the pitfalls with reporting accuracy in imbalanced class
settings, we report F1 and AUC metrics.

\subsubsection{Building interpretable models and extracting the predictive
value of features.}
\label{sec:predictors:methods:weights}

Given labels for each subreddit and a set of features associated with each
stage in its lifetime, we now seek to understand the predictive values of these
features. We achieve this in two steps: First, we build a machine learning
classifier model which uses these features to predict the labels associated with
each subreddit. Next, for high-performing classifiers, we analyze the weights
associated with each feature by the classifier. Our argument is that \emph{if
a classifier is able to achieve a reasonably good performance, then the features
it weighs heavily must have some predictive value}. Due to the need for
transparency in such models and administrative tasks, \emph{we focus solely on
interpretable models (logistic regressions, decision trees, and
random forests)}.

\para{Classifier model training, validation, and testing.} To evaluate the
performance of each classifier model we first split the samples in our dataset
with 80\% of each class randomly allocated for training and validation and the
remaining 20\% reserved for holdout testing. We then used five-fold
cross-validation to evaluate the classifier performance on the training and
validation dataset. We apply our oversampling and undersampling strategies
\emph{only on the training samples in each fold}. Finally, we evaluate the
classifier performance on the holdout set. In our results we report the mean
F1-score and AUC in the holdout samples.

\para{Interpreting models.} {Logistic Regression} models
a relationship between an outcome variable $y$ and a group of predictor
variables in terms of log odds. In order to interpret the model, we compute the
estimated weights for each feature and their corresponding odds ratio
\cite{ML-Interpret}. If the odds ratio for a feature ($f$) is $x$, it means
that a unit increase in $f$ changes the odds of our outcome variable $y$ by
a factor of $x$ when all other features remain the same. By calculating the
features with the highest odds ratios for different labels, we are able to
identify which features are the best predictors of problematic subreddits, as
later decided by Reddit administrators.
For our decision tree and random forest models, we find the importance of each
feature using Gini
Importance \cite{Breiman-Gini}. At a high-level the Gini importance counts the
number of times a feature is used as a splitting variable, in proportion with
the fraction of samples it splits. For random forests, the Gini importance is
averaged over all the constructed trees. We expect more important features to
have higher Gini importance scores. Unlike logistic regression interpretation,
a limitation here is that this metric only allows us to rank feature
importance, but not quantify the relative difference of their importance.

\subsection{Results}\label{sec:predictors:results}

%We now focus on measuring the effectiveness of using evolutionary features to
%predict which category (from dangerous/banned, hateful, and benign) subreddits
%belong to (\Cref{sec:predictors:results:accuracy}), which features are the best
%predictors of subreddit behavior (\Cref{sec:predictors:results:features}), and
%the impact of observation time on classifier accuracy
%(\Cref{sec:predictors:results:early}). Our results are summarized in 
%\Cref{tab:predictors:results}.

\subsubsection{Can we identify problematic subreddits by their evolutionary
features?} \label{sec:predictors:results:accuracy}

Column ``Total'' of \Cref{table:predictors:results:accuracy} shows how our
different explainable classifier models performed at classifying subreddits
into $\mathfrak{D}_I$ and $\mathfrak{D}_A$ when given access to all
evolutionary features of the subreddit, as would be available at the end of the
quarter Q4 of a subreddit's lifespan (which for $\mathfrak{D}_A$ subreddits is
the last month of data used in this study -- 04/2020). As we can see all our
models perform reasonably well, achieving F1-scores as high as 95\% on our
holdout set and a mean F1-score of up to 88\% in our five-fold
cross-validation experiments, regardless of whether models were built using
ADASYN oversampling or majority class undersampling. More interestingly, we
notice that our classifiers are able to achieve high F1-scores even as early as
after Q1 (between 91\% to 95\% AUC) with performance only increasing with
longer observations of a subreddit's evolution. These results indicate that, by
observing the evolutionary features described in
\Cref{sec:predictors:methods:features}, problematic subreddits can be
identified much earlier than they currently are. Further, the
performance of our interpretable classifiers are reasonable enough to warrant
their use to understand feature importance and the predictors of problematic
subreddits.

\begin{table}[t]
\small
  \resizebox{\columnwidth}{!}{
\begin{tabular}{c*{5}{>{$}c<{$}}}
\textbf{Model} & \multicolumn{4}{c}{\textbf{Classifier Score (\%AUC, \%F1-Negative, \%F1-Positive)}}  \\
\toprule
& \text{Q1}        & \text{Q1+Q2}       & \text{Q1+Q2+Q3}       & \text{Total}\\
\cmidrule{2-5}
RF-AS  & (93, 97, 68) & (95, 98, 78) & (96, 99, 84) & (97, 99, 88) \\
RF-US  & (91, 93, 65) & (91, 92, 68) & (92, 94, 74) & (94, 95, 80) \\
\cmidrule{1-1}
LR-AS  & (79, 79, 59) & (80, 81, 60) & (79, 83, 65) & (85, 92, 70) \\
LR-US  & (77, 76, 57) & (80, 82, 59) & (81, 83, 63) & (84, 86, 68) \\
\bottomrule
\\
\multicolumn{5}{l}{RF-AS = Random Forest with ADASYN sampling.}\\
\multicolumn{5}{l}{RF-US = Random Forest with random undersampling.} \\
\multicolumn{5}{l}{LR-AS = Logistic Regression with ADASYN sampling.} \\
\multicolumn{5}{l}{LR-US = Logistic Regression with random undersampling.}
\end{tabular}
}

\caption{Performance of our classifiers in predicting $\mathfrak{D}_I$ and
$\mathfrak{D}_A$ labels at the end of each quarter of a subreddit's lifespan.
% Q1 denotes that the classifier only had access to features from the first
% quarter of each subreddit's lifespan, Q1+Q2 denotes access to features from the
% first two quarters, Q1+Q2+Q3 denotes access to features from the first three
% quarters, and Total denotes access to features from the entire lifespan. 
Values denote area under the ROC, F1-score for non-problematic subreddits
  (F1-Negative), F1-scores for problematic subreddits (F1-Positive) on the
  holdout set.}
\label{table:predictors:results:accuracy}
\end{table}

% \subsubsection{How reactive is Reddit moderation to the media?}
% \label{sec:predictors:results:media}
% In \Cref{sec:introduction} we claimed that Reddit
% administrative actions were heavily driven by media criticism and therefore
% reactionary rather than proactive. To validate this claim, we removed all
% `mention'-related features from our classifier and compute its F1-score. We
% find that F1-score for the banned class drops to 76\%. While inspecting the
% confusion matrices obtained from our testing, we see that our classifiers had
% problems in distinguishing between subreddits with `banned' and `hate' labels
% (the F1-score for benign subreddits was 81\%). Including mentions-related
% features mitigates this problem. This result implies that: (1) while
% media-mentions are not predictive of hateful subreddits, they are predictive of
% banned subreddits (\ie these are predictive of admin reactions) (2) our
% classifier continues to have reasonable performance even when accounting for
% the absence of media-mention features, and (3) our classifier performs very
% well at filtering out benign subreddits -- thus showing its usefulness as a
% filtering tool for moderators.

\subsubsection{What features are most important for predicting the evolution
into problematic subreddits?} \label{sec:predictors:results:features}

% \begin{table}[t]
% \small
  % \begin{tabular}{lp{0.6in}p{0.6in}}
  % \toprule
    % {\textbf{Feature}}  & {\textbf{RF feature rank}} & {\textbf{LR log odds ratio}} \\
% %   {} & {\textbf{Feature rank}} & {\textbf{log odds ratio}} \\
    % \hline
    % \# users connected to \\
    % previously banned communities & 1   &   0.32\\
    % \# average percentage \\
     % of toxic comments score & 2 & 0.20 \\
    % \# negative mentions \\
     % in other communities& 3 & 0.24 \\
    % \# percentage of \\
    % comments removed & 4 & 0.18\\
    % \# negative mentions\\
     % in media outlets & 5 & 0.20\\
  % \bottomrule 
  % \end{tabular}
  % \caption{List of top-5 most predictive features for our random forest
  % (RF) and logistic regression (LR) classifiers. The top-4 features were
  % identical in both models, although with slightly different ordering of
  % importance.}
  % \label{tab:predictors:results:features} 
% \end{table}

Our random forest and logistic regression model were in agreement for the top-5
most important predictive features with slightly different ordering. We found that
the most predictive feature in both models is the `number of users who were
once active on banned communities'. This feature had the highest log-odds ratio
of .32 in our LR models while simultaneously ranking as the most important in
our RF models. This lends additional validity to our findings in
\Cref{sec:evolution}. Similarly, other features ranked in top-5 important
features by the random forest include average percentage of toxic comments (log
odds ratio: 0.20), negative mentions in other communities (log odds ratio:
0.24), percentage of comments removed (log odds ratio: 0.18), and negative
mentions in media outlets (log odds ratio: 0.20). These results suggest that
communities which entertain users with interactions in previously banned
communities have a significantly higher likelihood of becoming problematic as
well. Therefore, administrator interventions on problematic subreddits, rather
than the users of problematic subreddits, may not be the most effective
strategy for preventing the re-occurrence of problematic subreddits. 

\subsubsection{How well do explainable classifiers do as a real-world
administrative tool?}\label{sec:predictors:results:realworld}

Our previous results reflect our classifier performance over the entire dataset
of $\mathfrak{D}_I$ and $\mathfrak{D}_A$. We now seek to understand how well
our classifier would perform in a real-world deployment as a tool to aid Reddit
administration. We design a continuous learning \cite{chen2018lifelong}
experiment which emulates how Reddit administration would use our classifier --
with input from human administrators. A continuous learning framework is
important because content policies of Reddit have changed significantly over
time and leveraging a single snapshot of problematic subreddits does not allow
for our models to learn new patterns associated with subreddits which violate
newly added content guidelines -- \eg guidelines regarding the incitement of
violence were only added in 10/2017 therefore a model trained largely on prior
data would have little ability to identify problematic communities in this
category. 

\para{Experiment setup.} First, we begin by training our
RF-ADASYN model on data from 01/2018 to 06/2018. Next, we obtain the
subreddits identified as problematic by this model based on features obtained
from subreddits in 07/2018 only. From this list, we pay 
attention to three cases: (1) the identified subreddit was eventually banned or
quarantined by Reddit some time after 07/2018 (true positives), (2) the
identified subreddit was not banned or quarantined by Reddit (false positives),
and (3) ban or quarantine decisions that were made in 07/2018 that were not
identified as problematic by our model (false negatives). We then use the
false negatives as new training samples for the classifier so it may learn from
the human administrator's decisions while performing classifications for
subsequent months -- therefore mitigating the challenges associated with an
evolving content policy. This process is repeated for each month between
07/2018 and 04/2020. The false positives identified in each month are
indicative of problematic subreddits that have not yet been identified as such
by administrators and are potentially yet unknown to them while the true
positives serve as validation for our model's performance and allow us to
quantify the time advantage that proactive strategies yield -- \ie how much
earlier they are able to identify problematic communities.
We note that the
results obtained in this setup are not comparable to our previous results for
several reasons including: (1) different training periods -- our current setup
leverages only data from 01/2018 as opposed to our previous experiment which
included data from 2015/01, (2) different testing duration -- our test
features are obtained from just one month of the subreddit's lifespan as
opposed to an entire quarter, and (3) our model is retrained each month with
new administrator-identified problematic subreddits. It is precisely these
differences, which appear in a real world deployment, that warrant this
experiment.

\para{Results.} 
%\rnnote{Can we say something about mentions our TP/FN rate over time?}
Our model reported a total of 106 true positives and 26 false negatives. As
one might expect from the continuous learning setup, the false negative rate
decreased and the true positive rate increased over time. 
The 106 true positives included subreddits
    banned for toxicity (\eg \subreddit{TheRedPill} and
    \subreddit{The\_Donald}) and piracy (\eg \subreddit{soccerstreams}),
    amongst others. Our model identified them as problematic 9.3 months (mean)
    prior to their ban date by Reddit. 
Across the entire continuous learning
    experiment (until 04/2020), our model identified 43 subreddits as
    problematic that have not yet received administrative actions. These
    include \subreddit{KotakuInAction}, \subreddit{TumblrInAction},
    \subreddit{metacanada}, and \subreddit{MensRights}. We note that there have
    been several controversies and many reports of toxic behavior (\eg overt
    misoginy and racism) in these communities which support our model's
    decision. For example, \subreddit{KotakuInAction} was Reddit's
    primary pro-GamerGate community. In fact, in an effort to prevent the
    spread of toxicity, the subreddit was made private for a brief period
    during the peak of the movement. More recently, the community has expressed
    strong anti-transgender sentiment in the form of slurs and hate speech. Our
    classifier identified it as problematic based on features associated with
    vocabulary, and exceptionally high toxicity and negative media mentions (>2
    standard deviations from the mean). It remains unclear if the subreddits
    in our false-positives are receiving administrative attention from Reddit.
    In total our model suffered 26 false negatives.
    A large fraction of these were subreddits associated with eating disorders
    (\eg \subreddit{proED}, \subreddit{EDFood}, and \subreddit{thinspo})
    which were simultaneously banned due to their violation of a content policy
    regarding `encouraging self-harm'. We found that this was the first case of
    administrative action against such subreddits. As evidence of success in
    the continuous learning framework, we note that the subsequent quarantining
    of \subreddit{thinspocommunity} was correctly predicted by our model.
Taken together, our qualitative analysis suggests that our models are
effective and deployable in the real world as an administrative aid.

\subsubsection{Takeaway: Can evolution into problematic subreddits be
predicted?} \label{sec:predictors:results:takeaway}
Our results show that evolutionary features can be used to identify subreddits
likely to be problematic in the future. This finding validates hypothesis H2.
Our feature analysis which identified `number of users who were once active on
banned communities' as the most predictive suggests that interventions aimed at
users of banned communities might be an effective strategy to mitigate
problematic behavior. The explainable models also perform well in a continuous
learning real world deployment -- suggesting that they make effective
administrative aids.

%% file: relatedwork.tex
\section{Related Work}\label{sec:related}

We make contributions in two dimensions:
First, we perform measurements to understand how vocabulary and user bases of
online communities change over time (\Cref{sec:evolution}).
Second, we identify the predictors of problematic communities
(\Cref{sec:predictors}). In this section, we break down the related work in
each of these dimensions while highlighting their influence on our study.
% Specifically, in \Cref{sec:related:evolution} we explore related research
% seeking to understand how communities evolve and in
% \Cref{sec:related:incivility} we explore research aimed at measuring and
% mitigating incivility in online communities.

\para{Evolution of online communities.}
Studying behavioral patterns and evolution in online communities has been the
subject of several research efforts. These efforts can be taxonomized by
whether the goal is to understand evolution of interaction quantity or quality.
Research in characterizing interaction has generally focused on
understanding how the amount of interaction occurring in a community changes
over time and under different conditions. A general approach is to model
community interactions as a network graph where edges denote interactions (\eg
messages sent between two users) between nodes (\ie community members) and
track their evolution under different conditions. Especially relevant to our
work is research from Crandall \etal \cite{Crandall-KDD2008} which among other
results showed that \emph{interaction network related features are predictive
of future user behavior in topic-centered communities}. Researchers have also
tried to distinguish communities using interactive and linguistic features.
Mensah \etal \cite{mensah2020characterizing} observe growing and failing
subreddits in an attempt to distinguish their evolution using user interaction
and language patterns. Although their results show that there are no
significant differences in these features for growing and failing communities,
their results suggest the possibility of using interaction and linguistic
features as classifiers of other classes of communities. Several studies have
also investigated how specific user interactions are influenced by the age of a
community. Kiene \etal \cite{Kiene-CHI2016} showed that after a certain point
in the life-cycle of a community, large influxes of users had no impact on the
quality of discourse within the subreddit. \emph{These studies highlight the
need to consider age and stability of a community when predicting its
evolution}. Danescu \etal \cite{Danescu-WWW2013} found that linguistic features
in a community were constantly evolving and found that its newest members were
most likely to adapt their own linguistic features to those of the community.
Gazan \cite{Gazan-HICSS2009} found that, when communities stabilized, topics
tended to move away from topical and factual to personal and social. This
generally resulted in increased participation, often at the cost of conflict and
factionalism.
% Other research focused
% exclusively on understanding (rather than predicting) user interaction
% evolution in different communities. Kumar \etal \cite{Kumar-LM2010} studied the
% Flickr and Yahoo! 360 communities to understand the role of specific users in
% community growth. They found that a small number of key users are responsible
% for expanding a community and in the absence of these groups community growth
% stagnated. Ngamkajornwiwat \etal \cite{Ngamkajornwiwat-HICSS2008} focus on
% understanding how the network structure of online open source software
% development communities evolve over time. They found that relationships
% between ``\emph{core}'' and ``\emph{peripheral}'' developers generally weakened
% over time and contributions from the peripheral members tended to decrease.
% \emph{These results generally highlight the influence of a few community
% members on the direction and trends observed in a community}. 
% \para{Interaction quality.} Research in this category has generally focused on
% understanding how the nature of interactions (\ie its qualities) within
% a community change over time and under different conditions. 
% Garcia \etal \cite{Garcia-COSN2013}, in a post-mortem of Friendster, showed
% that {it is insufficient to consider only interaction network related
% features when modeling a community's \emph{resilience} to decline specifically
% highlighting the need to consider qualitative and external features}.
The importance of external events is highlighted by Zannettou \etal
\cite{Zannettou-IMC2018, Zannettou-IMC2017} who focused on the evolution of
memes and news sources within communities and uncovered their influence on
external communities. Focusing exclusively on Reddit, Mills \etal
\cite{Mills-AIS2018, Mills-SCSM2015} showed that, for \subreddit{The\_Donald}
and \subreddit{Sanders4President}, external events and their community
participation guidelines were largely responsible for their rise in popularity
and large influx of users. \emph{These studies highlight the need to consider
cross-community interactions and external events when considering evolution of
communities}.
%
%
% Focusing on the impact of \emph{status} in online communities,
% Bhalla \etal \cite{Bhalla-CMC2007} found that, over time, qualitative
% characteristics of a community began matching the characteristics of \emph{the
% elite}. Along similar lines, Gervais \cite{Gervais-PC2014} and Kwon \etal
% \cite{Kwon-IR2017} showed that exposure to incivility from political elites
% results in more  offensive rhetoric in online communities. 
In terms of methods,
we find most similarity between our approach and the work of Matias
\cite{Matias-CHI2016} which used a logistic regression model to
attribute weights to survey-derived features to uncover the factors associated
with moderators and subreddits participating in the Reddit-wide blackout of
2015 -- in protest of Reddit's administrative actions. They uncovered a strong
correlation between moderator participation in meta-reddit subreddits and
community participation in the protest. \emph{These findings further highlight
the important role played by a few key members (elites and moderators) in
a community}.  

\para{Predicting future community behavior.}
To maintain civil behavior in online communities timely identification and
removal and  violators is necessary as observed by Scrivens \etal
\cite{scrivens2020measuring}, they measure the evolution of radical posts
against particular vulnerable groups over-time. Their results show approval
(upvotes) shown towards hate speech increases gradually as users consistently
and frequently keep posting hate speech.These results corresponds with previous
works which show extremist communities polarizes opinion over time
\cite{caiani2015transnationalization, wojcieszak2010don, simi2015american}.
Furthermore, Seering \etal \cite{seering2019moderator} conduct a
semi-structured interview with moderators to find, among many other things, that
inconsistent moderation of communities lead to communities evolving chaotically
then requiring constant moderation and community policy updates.  Massanari
\cite{Massanari-NMS2017} conducted a qualitative analysis of the Reddit
communities at the center of the Fappening and Gamergate  controversies. The study highlights
how
the inaction of Reddit administrators and community moderators resulted in the
emergence of toxic technocultures and argues for the exploration of alternative
designs and moderation tools to combat the spread of such toxicity.
\emph{These findings highlight the importance and need of timely moderation and
intervention to maintain civil behavior}.
% research well being of moderators?
Research in automated moderation for online communities have been
mainly focused on content moderation at a `submission' or `content' level. Our
work aims to aid administrators perform moderation at a community level.
Chandrasekharan \etal \cite{chandrasekharan2019crossmod} created CrossMod,
a tool to aid Reddit moderators by detecting and moderating comments.
Similarly, Pavlopoulos \etal \cite{pavlopoulos2017deep} and Santos \etal
\cite{santos2021learning} developed and trained machine learning models to
detect violations by users on Wikipedia edits.

%% file: discussion.tex
\section{Conclusion}\label{sec:conclusions}

%Reddit has been the subject of many controversies on news and media outlets
%over the past few years. Its media coverage has facilitated a huge influx of
%users to this platform. This amount of growth in the user base has led to the
%formation of a variety of subreddits with different ideas and different sizes.
% In this paper, we investigated the feasibility of proactive moderation
% strategies for Reddit communities. 
%
%We find this phenomenon more frequently on dangerous subreddits. 

\para{Implications for other social platforms.} \revred{Due to the similarities
in community structure and platform designs, our methodologies to measure
evolution of communities and detect problematic communities has the potential
for application on other social media platforms such as Facebook.
For example, the concepts of communities, posts, comments, user migrations, and
administrator interventions have direct parallels with Facebook groups.
However, further investigation is needed to understand whether
similar evolutionary patterns can be exploited to develop moderation-aids on
such platforms where the online disinhibition effect
\cite{disinhibition-effect} might be weaker since users are not anonymous to
their communities and user accounts are required to reveal their real-world
identities \cite{fb-policy-anonymity}.} 

\para{The challenge of human-only moderation.}
We encounter cases in our study and anecdotally of communities that start off
as benign and evolve into communities that can later be categorized as
problematic. 
Currently, Reddit employs a small number of human administrators
\cite{We-are-the-nerds} to identify communities in violation of the Reddit
content policy and intervene to prevent future violations by those communities.
Due to the growing size of the platform, rather than seeking consistency in
policy enforcement, administrators are often only able to act in response to
user generated and media reports of egregious violations of the content policy.
Compounding their challenges, in \Cref{sec:evolution}, we showed that Reddit
communities, on average, are constantly evolving in both vocabulary and active
user bases. This implies the need for constant monitoring and attention, from
moderators and administrators, to proactively identify problematic communities
-- a prohibitively expensive proposition for human-only administration given
the large size of the platform. As regulations surrounding online social media
companies liability in content moderation (\eg \S230 of the US Penal Code) are
being re-evaluated world over, there is an urgency to develop tools to aid
human administrators perform such proactive identification and interventions at
scale.

% To maintain civil and constructive discourse and to moderate toxic and hateful
% content, Reddit currently uses a team of volunteer moderators and auto-moderator
% bots for each community. The responsibility of the moderators is to make sure
% the users adhere to the Reddit site rules and the community policies. Reddit
% also employs administrators who are tasked with moderating whole communities
% and are granted the authority to ban or quarantine toxic and hateful
% communities. 
% %
% The influx of users has made difficult for community moderators to
% enforce policies and for administrators to monitor devolving and dangerous
% communities. This suggests that Reddit’s current technique of human moderation
% is laborious and not feasible on either level.

% Predictive moderation
% Objective:
% The objective of this section is to first address the need and then present the
% predictive model as a solution to the aforementioned problem regarding the
% infeasibility of human moderators. We should also make clear that this system
% must not be used as an autonomous system.

% problem solution pattern?
% findings pattern?

% start with talking about moderation difficulty. Why we need this model ????

\para{Proactive moderation using predictive strategies.} %
%% shrink start.
%% no extreme words
%% easing to AI assistive
%% break yields sentence
% change conncetivity argument?
% known or hate to dangerous and hateful
% classify
% change the tools sentence
% remove proposal to histori...
% aid to focus moderation efforts.
% 
We observe that the evolutionary characteristics of problematic
subreddits is different from other subreddits. This yields opportunities for
providing machine-assisted human moderation.
We exploit these differences in evolutionary characteristics to build simple,
explainable, and accurate machine learning models to characterize the current
and predict the future behavior of different communities. 
%
%These models enables the prediction of (future misbehavior) with reasonably
%high accuracy. 
%
% Each of our classifiers show that structural properties of the community (\ie
% which other communities its users interact with) are the strongest predictors.
%
The accuracy of our predictions suggest that tools based on our approach and
features can be used to identify communities that are likely to exhibit
behavior similar to known problematic subreddits in the future.
Therefore, the output of these tools can be used by administrators to
proactively focus moderation efforts on a smaller set of communities. 
%
% In proposing these techniques we address the limitations of proactive automated
% moderation systems and underscore that this tool was not built for autonomous
% moderation as a replacement for human moderation. 
%
It is important to keep in mind that such proactive approaches, when used
autonomously, may have negative consequences. For example, there have been
reports of discrimination of LGBTQ content creators by YouTube's automated
content moderation system \cite{Youtube-demonetization} and, more critically,
strong racial profiling by autonomous predictive policing systems
\cite{Machine-bias}. Therefore, we only recommend using such tools to assist
human moderators by emphasizing which communities may require special attention.

%% file: paper.bbl
\begin{thebibliography}{47}
\providecommand{\natexlab}[1]{#1}
\providecommand{\url}[1]{\texttt{#1}}
\providecommand{\urlprefix}{URL }
\expandafter\ifx\csname urlstyle\endcsname\relax
  \providecommand{\doi}[1]{doi:\discretionary{}{}{}#1}\else
  \providecommand{\doi}{doi:\discretionary{}{}{}\begingroup
  \urlstyle{rm}\Url}\fi

\bibitem[{Angwin(2016)}]{Machine-bias}
Angwin, J. 2016.
\newblock Machine Bias.
\newblock Propublica.

\bibitem[{Baumgartner et~al.(2020)Baumgartner, Zannettou, Keegan, Squire, and
  Blackburn}]{pushshift}
Baumgartner, J.; Zannettou, S.; Keegan, B.; Squire, M.; and Blackburn, J. 2020.
\newblock The pushshift reddit dataset.
\newblock In \emph{Proceedings of the International AAAI Conference on Web \&
  Social Media}.

\bibitem[{Breiman(2017)}]{Breiman-Gini}
Breiman, L. 2017.
\newblock \emph{Classification and regression trees}.

\bibitem[{Caiani and Kr{\"o}ll(2015)}]{caiani2015transnationalization}
Caiani, M.; and Kr{\"o}ll, P. 2015.
\newblock The transnationalization of the extreme right and the use of the
  Internet.
\newblock \emph{International Journal of Comparative \& Applied Criminal
  Justice} .

\bibitem[{Chandrasekharan et~al.(2019)Chandrasekharan, Gandhi, Mustelier, and
  Gilbert}]{chandrasekharan2019crossmod}
Chandrasekharan, E.; Gandhi, C.; Mustelier, M.; and Gilbert, E. 2019.
\newblock Crossmod: A cross-community learning-based system to assist reddit
  moderators.
\newblock \emph{Proceedings of the Conference on Computer Supported Cooperative
  Work} .

\bibitem[{Chandrasekharan et~al.(2017)Chandrasekharan, Pavalanathan,
  Srinivasan, Glynn, Eisenstein, and Gilbert}]{Chandrasekharan-CSCW2017}
Chandrasekharan, E.; Pavalanathan, U.; Srinivasan, A.; Glynn, A.; Eisenstein,
  J.; and Gilbert, E. 2017.
\newblock You can't stay here: The efficacy of reddit's 2015 ban examined
  through hate speech.
\newblock \emph{Proceedings of the 2017 Conference on Computer Supported
  Cooperative Work} .

\bibitem[{Chawla et~al.(2002)Chawla, Bowyer, Hall, and
  Kegelmeyer}]{Chawla-IJCAI2002}
Chawla, N.~V.; Bowyer, K.~W.; Hall, L.~O.; and Kegelmeyer, W.~P. 2002.
\newblock SMOTE: synthetic minority over-sampling technique.
\newblock \emph{Journal of artificial intelligence research} .

\bibitem[{Chen and Liu(2018)}]{chen2018lifelong}
Chen, Z.; and Liu, B. 2018.
\newblock Lifelong machine learning.
\newblock \emph{Synthesis Lectures on AI \& Machine Learning} .

\bibitem[{Collins(2017)}]{HeatherHeyer-DailyBeast2017}
Collins, B. 2017.
\newblock Reddit Bans Forum Inciting `Physical Removal' of Democrats From
  Society.
\newblock The Daily Beast.

\bibitem[{Collins and Zadronzny(2018)}]{AlekMinassian-NBCNews2018}
Collins, B.; and Zadronzny, B. 2018.
\newblock After Toronto attack, online misogynists praise suspect as 'new
  saint'.
\newblock NBCNews.

\bibitem[{Crandall et~al.(2008)Crandall, Cosley, Huttenlocher, Kleinberg, and
  Suri}]{Crandall-KDD2008}
Crandall, D.; Cosley, D.; Huttenlocher, D.; Kleinberg, J.; and Suri, S. 2008.
\newblock Feedback Effects Between Similarity and Social Influence in Online
  Communities.
\newblock In \emph{Proceedings of the 14th ACM SIGKDD International Conference
  on Knowledge Discovery \& Data Mining}.

\bibitem[{Danescu-Niculescu-Mizil et~al.(2013)Danescu-Niculescu-Mizil, West,
  Jurafsky, Leskovec, and Potts}]{Danescu-WWW2013}
Danescu-Niculescu-Mizil, C.; West, R.; Jurafsky, D.; Leskovec, J.; and Potts,
  C. 2013.
\newblock No Country for Old Members: User Lifecycle and Linguistic Change in
  Online Communities.
\newblock In \emph{Proceedings of the 22nd International Conference on World
  Wide Web}.

\bibitem[{Dunning and Friedman(2014)}]{Dunning-2014}
Dunning, T.; and Friedman, E. 2014.
\newblock \emph{Practical Machine Learning: Innovations in Recommendation}.

\bibitem[{Facebook(2021)}]{fb-policy-anonymity}
Facebook. 2021.
\newblock What names are allowed on Facebook?

\bibitem[{Farokhmanesh(2018)}]{Youtube-demonetization}
Farokhmanesh, M. 2018.
\newblock YouTube is still restricting and demonetizing LGBT videos — and
  adding anti-LGBT ads to some.
\newblock The Verge.

\bibitem[{{Gazan}(2009)}]{Gazan-HICSS2009}
{Gazan}, R. 2009.
\newblock When Online Communities Become Self-Aware.
\newblock In \emph{2009 42nd Hawaii International Conference on System
  Sciences}.

\bibitem[{Han, Wang, and Mao(2005)}]{han2005borderline}
Han, H.; Wang, W.-Y.; and Mao, B.-H. 2005.
\newblock Borderline-SMOTE: a new over-sampling method in imbalanced data sets
  learning.
\newblock In \emph{International conference on intelligent computing}.

\bibitem[{He et~al.(2008)He, Bai, Garcia, and Li}]{he2008adasyn}
He, H.; Bai, Y.; Garcia, E.~A.; and Li, S. 2008.
\newblock ADASYN: Adaptive synthetic sampling approach for imbalanced learning.
\newblock In \emph{2008 IEEE international joint conference on neural
  networks}.

\bibitem[{Kiene, Monroy-Hern\'{a}ndez, and Hill(2016)}]{Kiene-CHI2016}
Kiene, C.; Monroy-Hern\'{a}ndez, A.; and Hill, B.~M. 2016.
\newblock Surviving an ``Eternal September'': How an Online Community Managed a
  Surge of Newcomers.
\newblock In \emph{Proceedings of the 2016 CHI Conference on Human Factors in
  Computing Systems}.

\bibitem[{Lagorio-Chafkin(2018)}]{We-are-the-nerds}
Lagorio-Chafkin, C. 2018.
\newblock \emph{We Are the Nerds: The Birth and Tumultuous Life of Reddit}.

\bibitem[{Liang et~al.(2016)Liang, Altosaar, Charlin, and
  Blei}]{Liang-RecSys2016}
Liang, D.; Altosaar, J.; Charlin, L.; and Blei, D.~M. 2016.
\newblock Factorization Meets the Item Embedding: Regularizing Matrix
  Factorization with Item Co-Occurrence.
\newblock In \emph{Proceedings of the 10th ACM Conference on Recommender
  Systems}.

\bibitem[{{Liu}, {Wu}, and {Zhou}(2009)}]{Liu-SMC2009}
{Liu}, X.; {Wu}, J.; and {Zhou}, Z. 2009.
\newblock Exploratory Undersampling for Class-Imbalance Learning.
\newblock \emph{IEEE Transactions on Systems, Man, and Cybernetics, Part B
  (Cybernetics)} .

\bibitem[{Marantz(2018)}]{Marantz-NY2018}
Marantz, A. 2018.
\newblock Reddit \& the Struggle to Detoxify the Internet.
\newblock \emph{The New Yorker} .

\bibitem[{Massanari(2017)}]{Massanari-NMS2017}
Massanari, A. 2017.
\newblock \# Gamergate \& The Fappening: How Reddit’s algorithm, governance,
  and culture support toxic technocultures.
\newblock \emph{New Media \& Society} .

\bibitem[{Matias(2016)}]{Matias-CHI2016}
Matias, J.~N. 2016.
\newblock Going dark: Social factors in collective action against platform
  operators in the Reddit blackout.
\newblock In \emph{Proceedings of the 2016 CHI conference on human factors in
  computing systems}.

\bibitem[{Mensah, Xiao, and Soundarajan(2020)}]{mensah2020characterizing}
Mensah, H.; Xiao, L.; and Soundarajan, S. 2020.
\newblock Characterizing the Evolution of Communities on Reddit.
\newblock In \emph{International Conference on Social Media \& Society}.

\bibitem[{Mills and Fish(2015)}]{Mills-SCSM2015}
Mills, R.; and Fish, A. 2015.
\newblock A computational study of how and why reddit. com was an effective
  platform in the campaign against sopa.
\newblock In \emph{International Conference on Social Computing \& Social
  Media}.

\bibitem[{Mills(2018)}]{Mills-AIS2018}
Mills, R.~A. 2018.
\newblock Pop-up political advocacy communities on Reddit. com:
  SandersForPresident and The Donald.
\newblock \emph{AI \& Society} .

\bibitem[{Molnar(2019)}]{ML-Interpret}
Molnar, C. 2019.
\newblock \emph{Interpretable Machine Learning: A Guide for Making Black Box
  Models Explainable}.

\bibitem[{Morse(2019)}]{RedditModeration-Mashable2019}
Morse, J. 2019.
\newblock Reddit waits until it's too late to ban violence-glorifying
  subreddits.
\newblock Mashable.

\bibitem[{Ohlheiser(2016)}]{Pizzagate-WaPo2016}
Ohlheiser, A. 2016.
\newblock Fearing yet another witch hunt, Reddit bans `Pizzagate'.
\newblock The Washington Post.

\bibitem[{Pavlopoulos, Malakasiotis, and
  Androutsopoulos(2017)}]{pavlopoulos2017deep}
Pavlopoulos, J.; Malakasiotis, P.; and Androutsopoulos, I. 2017.
\newblock Deep learning for user comment moderation.
\newblock \emph{arXiv preprint arXiv:1705.09993} .

\bibitem[{Pennebaker et~al.(2015)Pennebaker, Boyd, Jordan, and
  Blackburn}]{LIWC}
Pennebaker, J.~W.; Boyd, R.~L.; Jordan, K.; and Blackburn, K. 2015.
\newblock The development and psychometric properties of LIWC2015.
\newblock Technical report.

\bibitem[{Ribeiro et~al.(2020)Ribeiro, Jhaver, Zannettou, Blackburn,
  De~Cristofaro, Stringhini, and West}]{ribeiro2020does}
Ribeiro, M.~H.; Jhaver, S.; Zannettou, S.; Blackburn, J.; De~Cristofaro, E.;
  Stringhini, G.; and West, R. 2020.
\newblock Does Platform Migration Compromise Content Moderation? Evidence from
  r/The\_Donald \& r/Incels.
\newblock \emph{Proceedings of the 2013 Conference on Computer Supported
  Cooperative Work \& Social Computing} .

\bibitem[{Roberts(2014)}]{roberts2014behind}
Roberts, S.~T. 2014.
\newblock \emph{Behind the screen: The hidden digital labor of commercial
  content moderation}.
\newblock Ph.D. thesis, University of Illinois at Urbana-Champaign.

\bibitem[{Romano(2017)}]{RedditModeration-Vox2017}
Romano, A. 2017.
\newblock Why Reddit won't ban The\_Donald.
\newblock Vox.

\bibitem[{Santos, Osman, and Schorlemmer(2021)}]{santos2021learning}
Santos, T. F.~d.; Osman, N.; and Schorlemmer, M. 2021.
\newblock Learning for Detecting Norm Violation in Online Communities.
\newblock \emph{arXiv preprint arXiv:2104.14911} .

\bibitem[{Scrivens, Davies, and Frank(2020)}]{scrivens2020measuring}
Scrivens, R.; Davies, G.; and Frank, R. 2020.
\newblock Measuring the evolution of radical right-wing posting behaviors
  online.
\newblock \emph{Deviant Behavior} .

\bibitem[{Seering et~al.(2019)Seering, Wang, Yoon, and
  Kaufman}]{seering2019moderator}
Seering, J.; Wang, T.; Yoon, J.; and Kaufman, G. 2019.
\newblock Moderator engagement and community development in the age of
  algorithms.
\newblock \emph{New Media \& Society} .

\bibitem[{Simi and Futrell(2015)}]{simi2015american}
Simi, P.; and Futrell, R. 2015.
\newblock \emph{American Swastika: Inside the white power movement's hidden
  spaces of hate}.

\bibitem[{Suler(2004)}]{disinhibition-effect}
Suler, J. 2004.
\newblock The online disinhibition effect.
\newblock \emph{Cyberpsychology \& behavior} .

\bibitem[{Webber, Moffat, and Zobel(2010)}]{Webber-IS2010}
Webber, W.; Moffat, A.; and Zobel, J. 2010.
\newblock A Similarity Measure for Indefinite Rankings.
\newblock \emph{ACM Trans. Inf. Syst.} .

\bibitem[{Wohn(2019)}]{wohn2019volunteer}
Wohn, D.~Y. 2019.
\newblock Volunteer moderators in twitch micro communities: How they get
  involved, the roles they play, and the emotional labor they experience.
\newblock In \emph{Proceedings of the 2019 CHI conference}.

\bibitem[{Wojcieszak(2010)}]{wojcieszak2010don}
Wojcieszak, M. 2010.
\newblock ‘Don’t talk to me’: Effects of ideologically homogeneous online
  groups and politically dissimilar offline ties on extremism.
\newblock \emph{New Media \& Society} .

\bibitem[{Zannettou et~al.(2018)Zannettou, Caulfield, Blackburn, De~Cristofaro,
  Sirivianos, Stringhini, and Suarez-Tangil}]{Zannettou-IMC2018}
Zannettou, S.; Caulfield, T.; Blackburn, J.; De~Cristofaro, E.; Sirivianos, M.;
  Stringhini, G.; and Suarez-Tangil, G. 2018.
\newblock On the Origins of Memes by Means of Fringe Web Communities.
\newblock In \emph{Proceedings of the 2018 Internet Measurement Conference}.

\bibitem[{Zannettou et~al.(2017)Zannettou, Caulfield, De~Cristofaro,
  Kourtelris, Leontiadis, Sirivianos, Stringhini, and
  Blackburn}]{Zannettou-IMC2017}
Zannettou, S.; Caulfield, T.; De~Cristofaro, E.; Kourtelris, N.; Leontiadis,
  I.; Sirivianos, M.; Stringhini, G.; and Blackburn, J. 2017.
\newblock The Web Centipede: Understanding How Web Communities Influence Each
  Other Through the Lens of Mainstream \& Alternative News Sources.
\newblock In \emph{Proceedings of the 2017 Internet Measurement Conference}.

\bibitem[{Zhu, Xu, and Wu(2013)}]{Zhu-CDL2013}
Zhu, M.; Xu, C.; and Wu, Y.-F.~B. 2013.
\newblock IFME: Information Filtering by Multiple Examples with under-Sampling
  in a Digital Library Environment.
\newblock In \emph{Proceedings of the 13th ACM/IEEE-CS Joint Conference on
  Digital Libraries}.

\end{thebibliography}
